\documentclass[twocolumn,prb,floatfix,showpacs]{revtex4}
\usepackage{graphicx}

\newcommand{\crossprod}{\times}
\newcommand{\be}{\begin{equation}}
\newcommand{\ee}{\end{equation}}
\newcommand{\barr}{\begin{eqnarray}}
\newcommand{\earr}{\end{eqnarray}}
\newcommand{\breakeq}{\nonumber \\ &&}

\begin{document}

\title{Resonance shift effects in apertureless scanning
  near-field optical microscopy}

\author{J. A. Porto,$^1$ P. Johansson,$^2$ S. P. Apell,$^1$ 
and T. L\'opez-Rios$^3$}

\affiliation{
$^1$Department of Applied Physics, Chalmers University of Technology and
G\"oteborg University, S-41296  G\"oteborg, Sweden \\
$^2$Department of Natural Sciences, University of \"Orebro, S-701 82
\"Orebro, Sweden \\
$^3$Laboratoire d'Etudes des Propri\'et\'es Electroniques des Solides,
(LEPES/CNRS), BP 166, 38042 Grenoble Cedex 9, France}

\date{\today}

\begin{abstract}
We develop a theory to study apertureless scanning near-field optical
microscopy which takes into account retardation, higher multipoles of
the tip, and the multiple scattering between the tip and the surface. We
focus on metallic systems and discuss the implication of the formation
of tip-induced surface plasmon modes in the tip-surface system.  We
discuss the effects associated with the shift in energy of those modes
as a function of the tip-surface distance.  Both the local field and the
scattering cross section are enhanced when the tip approaches the
surface, but there is no general correspondence between the two
enhancements. 
\end{abstract}

\maketitle

\section{Introduction}
\label{sec:introduction}

Scanning near-field optical microscopy
(SNOM)~\cite{pohl84,lewis84,paesler} is a technique that allows to
obtain optical images of objects with subwavelength resolution.  In the
pioneering experiments,~\cite{pohl84,lewis84} subwavelength optical
apertures where used to illuminate the sample and obtain subwavelength
images. Nevertheless, it is also possible to perform near-field optical
imaging without aperture probes.  This is the case of  apertureless
SNOM,~\cite{fischer,specht,zenhausern,inouye,gleyzes} which has
attracted considerable attention in recent years, allowing interesting
experiments such as near-field fluorescence imaging~\cite{sanchez} and
near-field vibrational absorption detection.~\cite{knoll} In
apertureless SNOM, light is focused at the tip of a scanning probe
microscope (SPM) and the field enhancement near the tip is used to
obtain the subwavelength imaging. Apertureless SNOM allows to perform
simultaneously near-field optical imaging and other types of local probe
microscopies, such as atomic force microscopy (AFM) or scanning
tunneling microscopy (STM). In addition, it has provided higher
resolution than other SNOM techniques.  A key issue in the apertureless
SNOM technique is the tip-sample coupling, and how the  electromagnetic
resonances at the cavity formed by the tip and the surface may affect
the corresponding scattered field.

There has been an extensive theoretical research about
SNOM,~\cite{girard96,greffet} mostly related to imaging with aperture
probes.  With regard to apertureless SNOM, it has been studied
theoretically considering different models and approximations such as
electric dipole for the tip,~\cite{vigoureux,girard92} the electrostatic
approximation,~\cite{denk} or the passive probe
approximation.~\cite{vanlabeke} It has also been studied in two
dimensions.~\cite{madrazo} There have been detailed numerical
calculations of the near field associated with apertureless
probes,~\cite{martin,novotny97,furukawa} and the field enhancement has
been discussed in the context of apertureless SNOM.~\cite{zayats} In
addition, the role of tip geometry,\cite{aigouy,aizpurua,porto00} the
influence of tip modulation,~\cite{walford01} and the implications for
magneto-optical near-field imaging~\cite{johansson01,walford02} have
been studied.  Nevertheless, there is a need for clarification of how
the tip-sample coupling affects the signal detected and the local-field
enhancement.  In particular, the question concerning the link between
electromagnetic resonances at the cavity formed by the tip and the
surface and the corresponding scattered field needs to be addressed.
These resonances concern free electron metals for which surface plasmons
may exist. In the simple case of an isolated particle it is well known
that the scattering and extinction cross sections have resonance maxima
at about the same photon energy. This is obvious from the fact that
large charges at the surface give rise  to  large  dipoles  and
consequently large radiation.  If we instead consider two interacting
particles the situation is not so simple. In this  case  there  is  not
a  direct correlation between near-field and far-field
resonances.\cite{aravind,metiu}

This problem is of the highest relevance concerning the emerging field
of optical spectroscopy with an apertureless SNOM, which comprises areas
such as near-field Raman scattering and luminescence. The small number
of molecules between the tip and the surface leads to extremely small
signals that are very difficult to measure. One possible way to overcome
this problem is to take advantage of the resonance that may exist at the
cavity between the tip and the surface and the extremely high fields
associated.~\cite{rendell} In this context, a key question is what is
the signature of this type of resonances on the far field and on the
optical images.

In this article we develop a theory which allows to perform a detailed
study of the tip-sample interaction in apertureless SNOM.  We model the
tip by a sphere of finite size, whose radius may correspond roughly to
the radius of curvature of the real tip.  The main features of the
theory are: (i) retardation, (ii) higher multipoles of the sphere, and
(iii) the multiple scattering between the sphere and the surface are
taken into account.~\cite{johansson01} We use this theory to analyze the
effects associated with the shift in energy of the surface-plasmon
resonances of the tip-sample system as the tip approaches the surface.
Those resonances produce an enhancement in the near field and in the far
field as well, although there is not always an exact correspondence
between the two enhancements.  We will also discuss the convergence of
the results with the number of multipoles taken into account in the
calculation. As a general result, a high number of multipoles should be
retained to arrive at convergence. 

The paper is organized as follows. In Sec.~\ref{sec:method} we highlight
the main points of the theoretical framework used for the calculations.
In Sec.~\ref{sec:results} we discuss the numerical results obtained for
the near-field enhancement and scattering cross sections as a function
of different parameters of the system.  In Sec.~\ref{sec:conclusions},
we present the concluding remarks.

\section{Method}
\label{sec:method}

\begin{figure}[htb]
\includegraphics[angle=0, width=8.5cm]{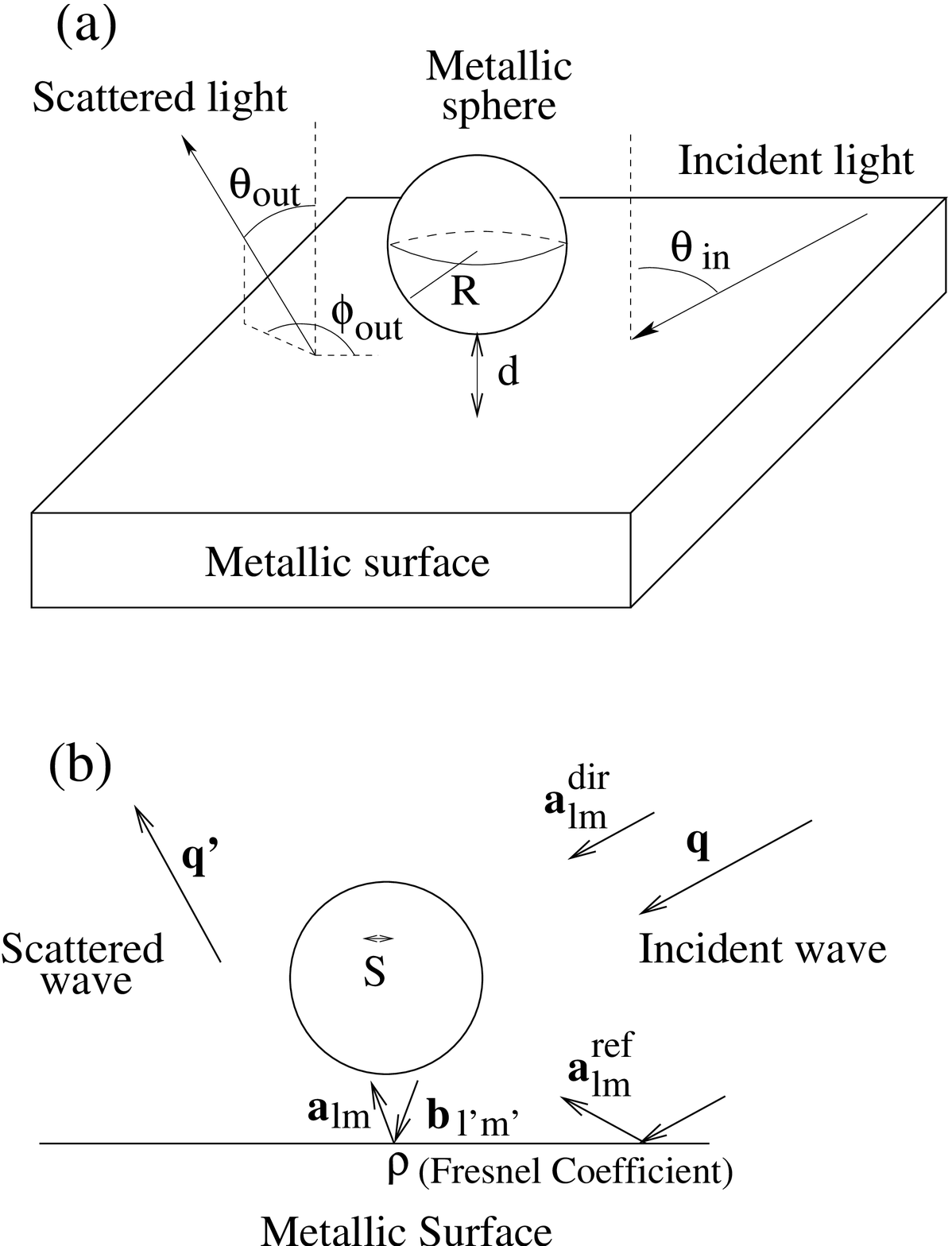}
\vspace{0.5 cm}
\caption{ 
   (a) Model geometry used in the calculation and (b) illustration of
   the scheme used in the calculation
}
\label{fig:scheme}
\end{figure}

In this section we present the main features of the theoretical
formalism employed.~\cite{johansson01} Figure~\ref{fig:scheme}(a) shows
the system under study.  A flat metallic sample, filling the half-space
$z<z_0=-(R+d)$, is probed by a metallic tip, modeled by a sphere with
radius $R$.  The distance between the sphere and the surface is $d$, as
shown in Fig.~\ref{fig:scheme}(a).

We are interested in studying how this system scatters the incident
light and which local electromagnetic fields are built up in the
process.  For that purpose, we consider a plane wave incident on this
system with wave vector  ${\bf q}$, which can be written as (an
$e^{-i\omega t}$ factor is omitted in fields and other quantities in
what follows)
\be
  {\bf E}_{in} ({\bf r})= \left\{ E^{(s)} \hat{\bf s}
 + E^{(p)} \hat{\bf p}  
             \right\}
  e^{i{\bf q}\cdot{\bf r}}.
\label{incoming}
\ee
For a wave vector ${\bf q}$, the polarization vectors for s and p
polarization are, respectively,
$\hat{\bf s}= \hat{\bf z}\crossprod \hat{\bf q}_{\|}$, and
$ \hat{\bf p}= 
- \hat{\bf q} \crossprod (\hat{\bf z}\crossprod \hat{\bf q}_{\|}),$ 
where ${\bf q}_{\|}$ is the projection of ${\bf q}$ in the surface plane.
In the present work we will mainly focus on the p-polarization. 

In order to tackle the scattering off the sphere-surface system, the
electromagnetic field inside and just outside the sphere is expanded in
terms of the electric (E) and magnetic (M) multipoles.~\cite{jackson}
The electric field inside the sphere can be expressed as 
\be 
 {\bf E}=\sum _{lm} k \, c_{lm}^{(M)}\,  j_l(k_T r){\bf X}_{lm}+
  \frac{i}{\epsilon_T}\nabla\crossprod
   \left[c_{lm}^{(E)} j_l(k_T r){\bf X}_{lm}\right]. 
   \ee 
The electric field just outside the sphere can be written as
the linear combination of incident and outgoing waves from the sphere
\barr 
 {\bf E}= &&\sum _{lm}
  k \, [a_{lm}^{(M)}\,j_l(k r) + b_{lm}^{(M)} h_l(kr)]\, {\bf X}_{lm}+
\breakeq
 + i\,\nabla\crossprod
 \left\{[a_{lm}^{(E)} j_l(kr) + b_{lm}^{(E)} h_l(kr)]\, {\bf X}_{lm}\right\},
\label{Eseries}
\earr
where ${\bf X}_{lm}$ is the vector spherical harmonics, $j_l$ denotes
the spherical Bessel functions related to incoming waves, and $h_l$ are
the spherical Hankel functions related to outgoing waves. Equivalent
expressions can be written for the magnetic field.

By imposing the continuity of the tangential components of {\bf E} and
{\bf H} and the normal components of {\bf B} and {\bf D} at the sphere
surface, it is possible to relate the coefficient of the outgoing wave
to the incoming wave for each electric and magnetic multiple:
$$
s_l^{(E)} = \frac{b_{lm}^{(E)}}{a_{lm}^{(E)}},\ \ {\rm and} \ \ 
s_l^{(M)} = \frac{b_{lm}^{(M)}}{a_{lm}^{(M)}},
$$
where $s_l^{(E)}$ and $s_l^{(M)}$ are sphere response functions.
Due to the symmetry of the sphere, the ``quantum'' number $l$ is 
conserved in a scattering event
and the response functions are independent of $m$.
These expressions can be rewritten in a matrix form
\be
 \vec{b} = \tensor{s} \vec{a},
\label{rel}
\ee 
where $\tensor{s}$ is a diagonal tensor that relates the incoming wave
coefficients, $\vec{a}$, to the outgoing waves, $\vec{b}$.

In the situation under study, the field impinging on the sphere consists
of three contributions [see Fig.~\ref{fig:scheme}(b)],
\be
 \vec{a} = \vec{a}^{\rm dir} + \vec{a}^{\rm ref} + \tensor{N} \vec{a}.
\label{asyst}
\ee
$\vec{a}^{\rm dir}$ corresponds to the amplitude of the field from the
original incident  wave, its value can be obtained by expanding the
incident plane wave in the multipoles of the sphere, while 
$\vec{a}^{\rm ref}$  represents the amplitude of the incident wave
reflected once from the sample surface. Its value is obtained by
expanding the reflected plane wave from the surface in the multipoles of
the sphere.  Finally, $\tensor{N}\vec{a}$ represents waves scattered
from the sphere that return to it after reflection from the sample.
This term contains the exact field described by $\vec{a}$, therefore it
accounts for all multiple scattering events in which waves are scattered
between the sphere and the sample an arbitrary number of times.
$\tensor{N}$ is a tensor in multipole space. It indicates how a
spherical wave with a given angular momenta l and m is reflected from
the surface and returns to the sphere. In this reflection process the
quantum number $l$ is not necessarily conserved but, thanks to the
cylindrical symmetry, $m$ is. In the coupling between
the sphere and the sample, both propagating and evanescent waves are
taken into account. 

Retaining a finite number of multipoles,  Eq.~(\ref{asyst}) can be 
solved by a matrix inversion
\be 
 \vec{a} = \left[\tensor{1} - \tensor{N} \right]^{-1} \left(
  \vec{a}^{\rm dir} + \vec{a}^{\rm ref} 
  \right).
\label{asolveq}
\ee

Now, from Eq.~(\ref{rel}) we can obtain the coefficients for the
outgoing waves from the sphere. From those coefficients it is possible
to calculate the radiated field from the sphere-surface system in a
particular direction given by the wave vector  ${\bf q}'$,
\be
{\bf E}_{\rm rad}=
\left\{ E_{\rm rad}^{(s)} \hat{\bf s}'
+ E_{\rm rad}^{(p)}
\hat{\bf p}' 
\right\}
e^{i{\bf q}'\cdot{\bf r}}.
\label{radiated}
\ee

If we consider a direction other than the specular direction, the
radiated field comprises two contributions:
\be
{\bf E}_{\rm rad}= {\bf E}_{\rm rad,sphere} + {\bf E}_{\rm rad,surface}, 
\ee
where ${\bf E}_{\rm rad,sphere}$ represents the field radiated directly
by the sphere, and ${\bf E}_{\rm rad,surface}$ corresponds to waves
having a final scattering event from the surface.

\section{Results and discussion}
\label{sec:results}

\begin{figure}[b]
\includegraphics[angle=0, width=8cm]{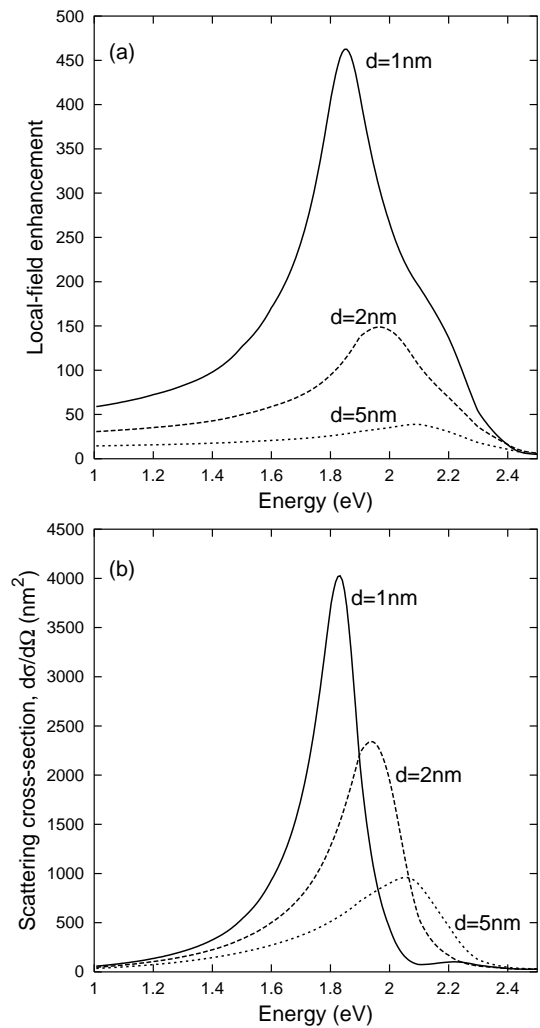}
\caption{ 
   (a) Near-field enhancement just underneath the sphere and (b)
   differential scattering cross section as a function of the energy of
   the incident wave for a gold sphere of radius R=40~nm, and for
   different values of the distance between the tip and the surface,
   $d$.  The angle of incidence is $\theta_{\rm in}=57^{\rm o}$ and the
   angles of observation are $\theta_{\rm out}=57^{\rm o}$ and
   $\phi_{\rm out}=90^{\rm o}$.
}
\label{fig:R40nm}
\end{figure}

In this section, we study the system formed by a gold sphere and a gold
surface by means of the theoretical method described above.  The
dielectric function of gold is described using the tables reported in
Ref.~\onlinecite{palik}. We will consider an incident p-polarized wave
with angle of incidence  $\theta_{in} = 57^o$.  We will mainly
concentrate on
the effects associated with the formation of resonances of the
tip-surface coupled system.  In what follows, we will mainly focus on
the analysis of the local field enhancement and the scattering cross
section for scattered p-polarized light. Results will be shown for the
observation angles $\theta_{\rm out}=57^{\rm o}$ and 
$\phi_{\rm out}=90^{\rm o}$.  It should be noticed that similar effects
than the ones discussed in the present work can be seen for other angles
of observation and incidence. 

\begin{figure}[b]
\includegraphics[angle=0, width=8cm]{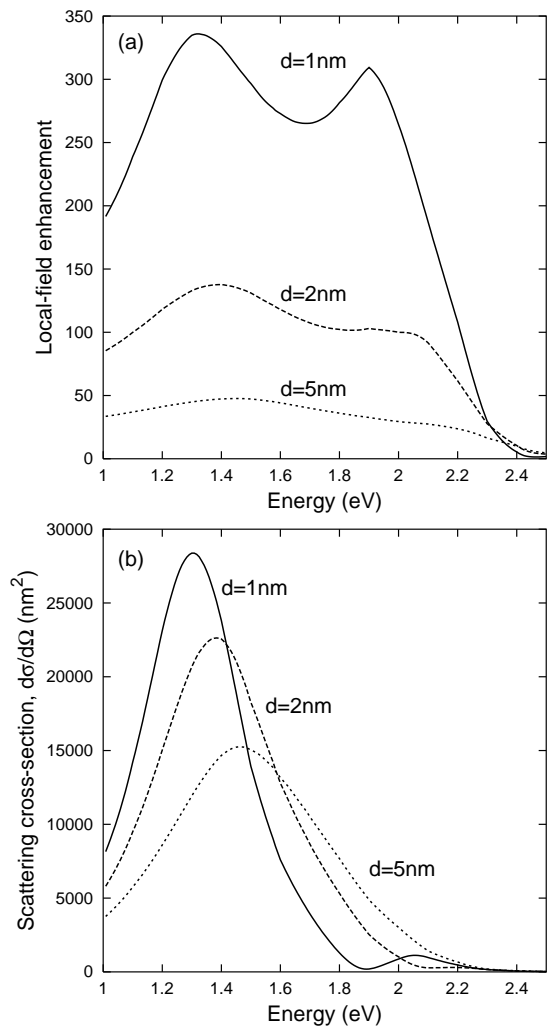}
\caption{
   (a) Near-field enhancement just underneath the sphere and (b)
   differential scattering cross section as a function of the energy of
   the incident wave for a gold sphere of radius R=80~nm, and for
   different values of the distance between the tip and the surface,
   $d$.  The angle of incidence is $\theta_{\rm in}=57^{\rm o}$ and the
   angles of observation are $\theta_{\rm out}=57^{\rm o}$ and
   $\phi_{\rm out}=90^{\rm o}$.
}
\label{fig:R80nm}
\end{figure}

Figure~\ref{fig:R40nm}(a) shows the local-field enhancement below the
sphere as a function of energy for different values of the tip-sample
distance and a sphere radius of 40~nm. The peak that is built up as the
sphere approaches the surface is due to a surface plasmon of the
coupled system formed by the sphere and the surface.~\cite{rendell} This
maximum experiences a redshift as the tip approaches the surface. The
scattered light from the system shows a similar peak, as illustrated by
Fig.~\ref{fig:R40nm}(b). 

Figure~\ref{fig:R80nm}(a) shows the local-field enhancement when a
larger sphere (radius of 80~nm) is considered. The first peak has moved
to lower energy (compared to the previous case with smaller radius)
and a new peak appears at higher energy, but the field enhancement does
not vary substantially over a wide range of energies. 
The scattering cross section has a qualitatively different
behavior [see Fig.~\ref{fig:R80nm}(b)].  The first peak appears also in
the scattering cross section.  By contrast, for higher energies, there
is no significant correspondence between the scattering cross section
and the local field enhancement. 

\begin{figure}[b]
\includegraphics[angle=0, width=8cm]{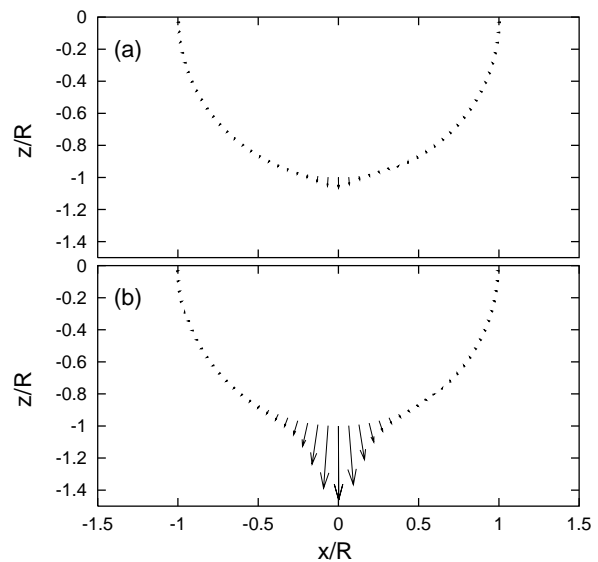}
\caption{
   Real part (a) and imaginary part (b) of the electric field on the
   sphere surface for $R=80~nm$, $E=1.32~eV$ and $d=1~nm$, which
   corresponds to the first maximum of the near-field enhancement in
   Fig.~\protect\ref{fig:R80nm}(a). The displayed fields have been
   calculated just outside the surface of the sphere in a vertical plane
   that contains the incident wave vector ${\bf q}$. The absolute
   magnitude of the depicted fields is about 325 times stronger than the
   incident field as can be seen in Fig.~\protect\ref{fig:R80nm}.
}
\label{fig:fm80}
\end{figure}

\begin{figure}[b]
\includegraphics[angle=0, width=8cm]{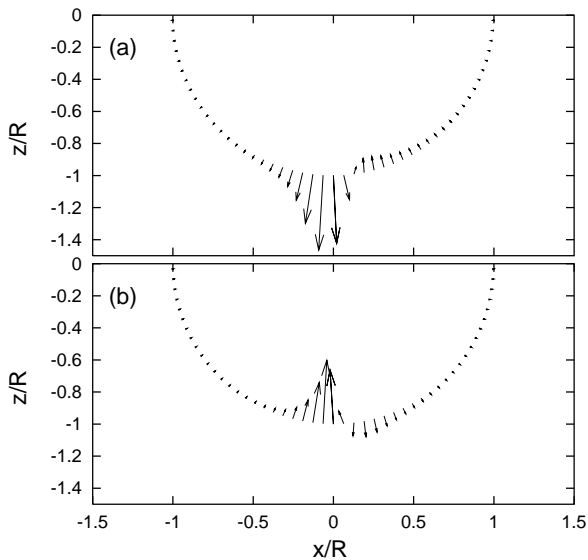}
\caption{
   Real part (a) and imaginary part (b) of the electric field on the
   sphere surface for $R=80~nm$, $E=1.9~eV$ and $d=1~nm$, which
   corresponds to the second maximum (with a magnitude $\sim$ 300 times
   that of the incident field) of the near-field enhancement
   in Fig.~\protect\ref{fig:R80nm}(a).  
}
\label{fig:hm80}
\end{figure}

In order to analyze the different features in
Figs.~\ref{fig:R40nm}~and~\ref{fig:R80nm}, we study the distribution of
the electric field near the sphere.  Figure~\ref{fig:fm80} represents
the electric field around the 80~nm-radius sphere for the first field
enhancement maximum in Fig.~\ref{fig:R80nm}(a). This figure, as well as
other vector plots shown below, displays
the electric field just outside the sphere in a vertical plane that
also contains the incident wave vector ${\bf q}$.  
The fields shown in Fig.~\ref{fig:fm80} corresponds
to the fundamental resonant surface-plasmon mode of the tip-surface
system.  This fundamental mode is also responsible for the peak in
Fig.~\ref{fig:R40nm}(a), where a smaller radius is considered. The local
field in that case shows a very similar pattern. However, for the second
peak in Fig~\ref{fig:R80nm}(a), the field pattern is significantly
different, as can be seen in Fig.~\ref{fig:hm80}. The local field shows
extra features that correspond to a higher mode of the tip-surface
system.  The dipolar character of the fundamental mode makes it fairly
efficient in emitting light, as shown by the large scattering cross
section associated with it (see Fig.~\ref{fig:R80nm}(b)). On the
contrary, the higher mode has a small scattering cross section
associated, which can be explained by the more complex distribution of
the local field for this mode.  

\begin{figure}[h]
\includegraphics[angle=0, width=8cm]{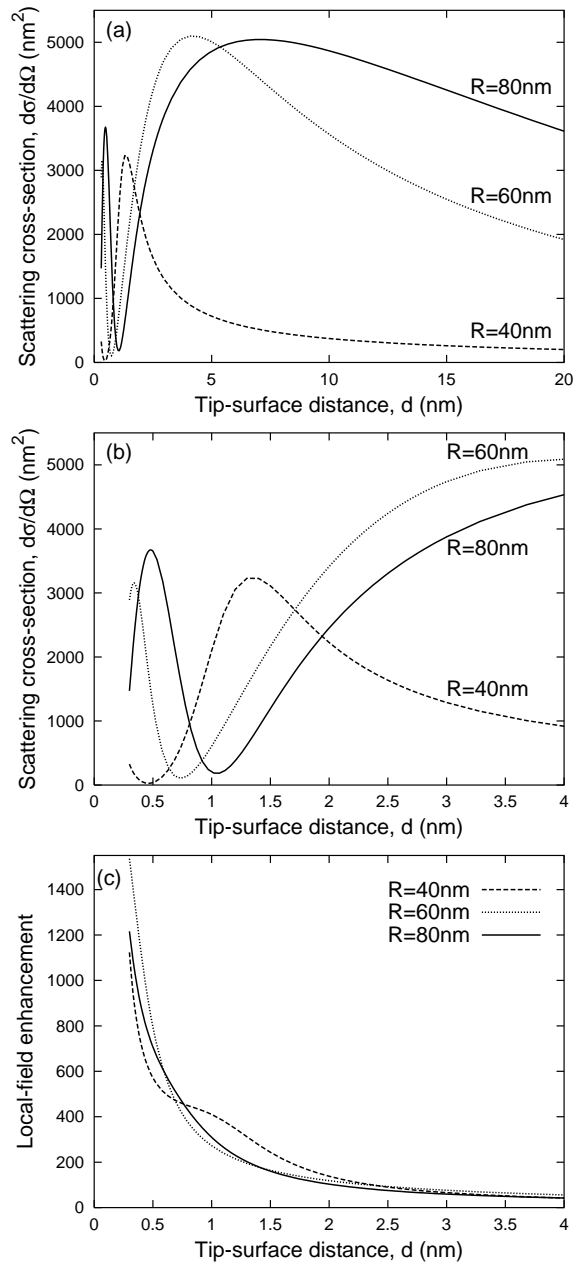}
\caption{
   Differential scattering cross section (a)  and (b), and near-field
   enhancement just underneath the sphere (c), as a function of the
   tip-surface distance $d$, for a photon energy of 1.9~eV and sphere
   radius $R$ of 40~nm, 60~nm, and 80~nm.  The angle of incidence is
   $\theta_{\rm in}=57^{\rm o}$ and the angles of observation are
   $\theta_{\rm out}=57^{\rm o}$ and $\phi_{\rm out}=90^{\rm o}$.
}
\label{fig:1.9eV}
\end{figure}

The formation of such resonances and their energy shifts have
significant consequences on the ``approach curves'', {\it i.e.}\ when
varying the tip-surface distance for fixed values of the photon energy.
Figure~\ref{fig:1.9eV} shows the local-field enhancement and the
scattering cross section for a photon energy of 1.9~eV and different
values of the radius of the sphere. The local-field enhancement
increases when the tip approaches the surface for the range of
tip-sample distance studied and for all the sphere radii under study
[Fig.~\ref{fig:1.9eV}(c)].  (We only show results for tip-surface
distances larger than 0.3~nm, since for very short distances the
description of the system with a macroscopic dielectric function is no
longer justified.~\cite{johansson98}) 

Figure~\ref{fig:1.9eV}(a) illustrates the behavior of the differential
scattering cross section as a function of tip-surface distance. It
presents a maximum. It is worth noticing that this maximum does not
imply a maximum of the local field enhancement, which keeps increasing
when the sphere approaches the surface.  The corresponding feature in
the local field enhancement is a shoulder, rather than a maximum, as can
be seen in Fig~\ref{fig:1.9eV}(c) for sphere radius R=40~nm.  The
tip-surface distance at which the maximum of the scattered field appears
increases with the radius of the sphere and the peak becomes wider. In
addition, some extra structure appears at very short distances [see
fig.~\ref{fig:1.9eV}(b)], especially for the larger radii. We will
analyze the physical origin of the different features by looking in
detail at the local field distribution. 

\begin{figure}[htb]
\includegraphics[angle=0, width=8cm]{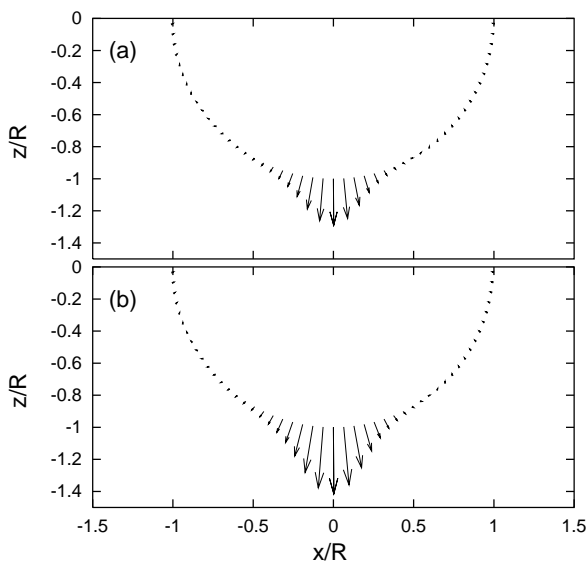}
\caption{
   Real part (a) and imaginary part (b) of the electric field on the
   sphere surface for $R=40~nm$, $E=1.9~eV$ and $d=1.23~nm$, which
   corresponds to a peak of scattering cross section in
   Fig.~\protect\ref{fig:1.9eV}(b). The local field is about 400 times
   stronger than the incident field in this case as can be seen in
   Fig.~\protect\ref{fig:1.9eV}(c).
}
\label{fig:fm1.9eV}
\end{figure}

\begin{figure}[htb]
\includegraphics[angle=0, width=8cm]{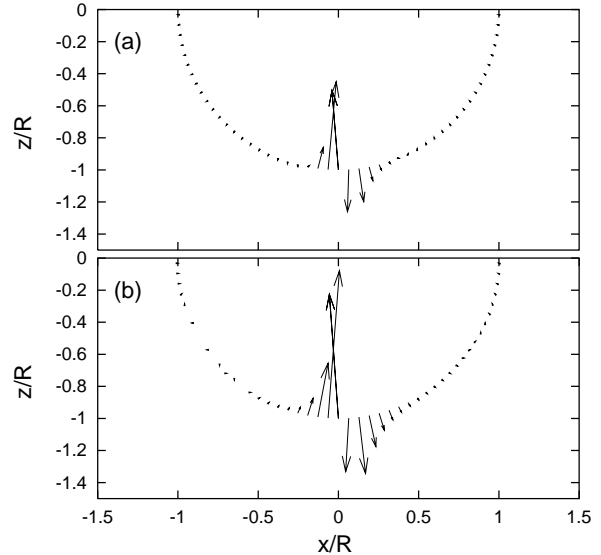}
\caption{
   Real part (a) and imaginary part (b) of the electric field on the
   sphere surface for $R=80~nm$, $E=1.9~eV$ and $d=0.48~nm$, which
   corresponds to a peak of scattering cross section in
   Fig.~\protect\ref{fig:1.9eV}(b). From Fig.~\protect\ref{fig:1.9eV}(c)
   we see that the local field in this case is about 1000 times stronger
   than the incident field.
}
\label{fig:hm1.9eV}
\end{figure}

Figure~\ref{fig:fm1.9eV} displays the local field around the sphere for
the maximum of the scattering cross section for a sphere with radius
40~nm.  This figure shows that the peak corresponds to the fundamental
surface plasmon mode of the tip-surface system.  The approach curves for
small spheres were previously calculated by Girard,\cite{girard92} who
modeled the sphere by an electric dipole and obtained peaks that can be
related to the fundamental surface plasmon mode.  For the larger radii
under study, the fundamental mode resonance is already shifted to
energies lower than 1.9~eV, the photon energy considered in
Fig.~\ref{fig:1.9eV}.  The behavior of the scattering cross section as a
function of tip-surface distance is in this case controlled by the tail
of the resonance peaks shown in Fig.\ \ref{fig:R80nm}. This explains
why the peaks in the scattering cross section {\em as a function of
tip-sample distance} displayed in Fig.~\ref{fig:1.9eV}(a) and
calculated for larger sphere radii (60 nm and 80 nm) are much
wider than the one calculated for the 40-nm sphere.
At a much smaller tip-surface
distance, i.e.\ about 0.5~nm, a narrow peak is present for the $R$=80~nm
sphere [Fig.~\ref{fig:1.9eV}(b)]. The local field distribution displayed
in Fig.~\ref{fig:hm1.9eV}, with its characteristic node, shows that this
peak is associated with the higher mode resonance of the tip-surface
system.  The tip-sample distance at which this higher-mode resonance
appears increases with increasing sphere radius and photon energy.

When studying the light scattered by an isolated sphere one finds that
for the larger sphere ($R$=80~nm) there is a Mie
resonance at energies near 1.9~eV.  When considering the sphere-surface
system,  the Mie resonance of the isolated sphere implies that, for the
80~nm-radius sphere, the amount of s-polarized scattered light becomes
larger than the p-polarized light.  Nevertheless, the amount of
s-polarized scattered light is almost independent of the tip-sample
distance, except for distances shorter than 1~nm, where some features
appear that could be related to the formation of the higher resonant
mode discussed above. 

\begin{figure}[tb]
\includegraphics[angle=0, width=8cm]{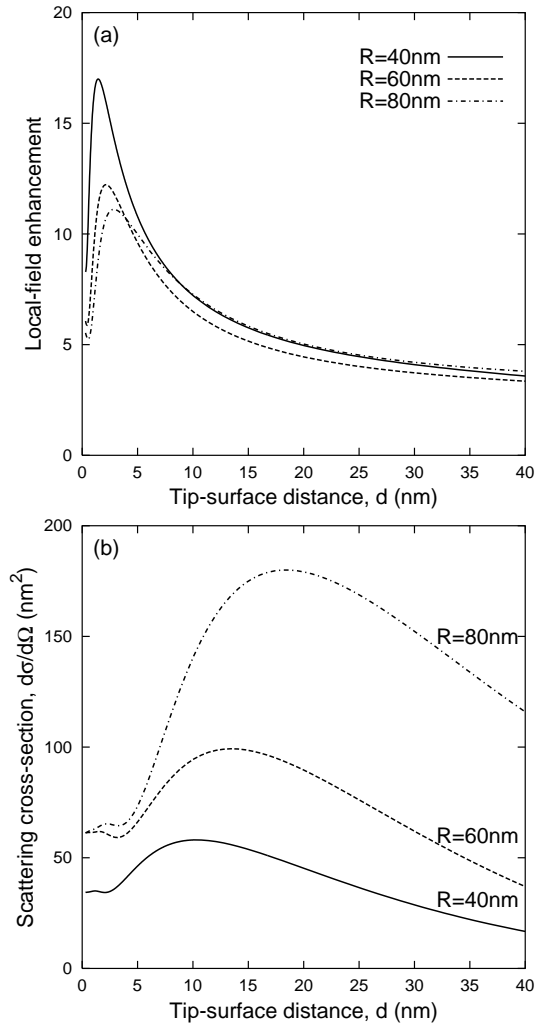}
\caption{
   (a) Near-field enhancement just underneath the sphere, and (b)
   differential scattering cross section as a function of the
   tip-surface distance $d$, for a photon energy of 2.4~eV and  sphere
   radii of 40~nm, 60~nm, and 80~nm, respectively.  The angle of
   incidence is $\theta_{\rm in}=57^{\rm o}$ and the angles of
   observation are $\theta_{\rm out}=57^{\rm o}$ and 
   $\phi_{\rm out}=90^{\rm o}$.
}
\label{fig:2.4eV}
\end{figure}

It is also interesting to analyze the approach curve for photon energies
with no significant resonance effects.  Figure~\ref{fig:2.4eV}(a) shows
the local-field enhancement for a photon energy of 2.4~eV\@.  In this
``off-resonance'' situation, the local-field enhancement presents a
maximum as a function of the tip-sample distance. This maximum appears
between d=1~nm and d=3~nm for the sphere radii studied.  The maximum of
the local field enhancement manifests itself also in the scattering
cross section through a very small peak at similar tip-surface distance
[see Fig.~\ref{fig:2.4eV}(b)].  However, the scattering cross-section
maximum appears at much larger tip-surface distances, between d=10~nm
and d=20~nm.

In order to gain physical insight about the origin of the different
maxima in the off-resonance situation ($E$=2.4~eV), we study the
convergence with the number of multipoles taken into account in the
calculation.  We analyze the case of a sphere radius of 40~nm, for which
the maxima of enhancement and scattered amplitude appear at well
separated values of tip-surface distance.  The local-field enhancement
requires a high number of multipoles to converge, in this particular
case it was necessary to consider multipoles of order $l=20$ to arrive
at a satisfactory convergence. Differently, convergence
for the scattering cross section is achieved with a low number of multipoles.
Retaining multipoles up to $l=2$, {\it i.e.}\ dipole and quadrupole
modes of the sphere, we obtain most of the features of the scattering
cross section, in particular, its main maximum.  This fast convergence
implies that a simple model for the sphere-surface system can be
adequate for the estimation of the scattering off the system when no
resonant behavior is involved. In particular, the often used model consisting
of a dipole representing the sphere and its image dipole that together
have a quadrupole character might be suitable in this situation.  If we
are interested in the local-field enhancement, even in this off
resonance situation, we need to consider a  more sophisticated model
including higher multipoles of the sphere.

On the other hand, for a photon energy of 1.9~eV (the ``on-resonance''
situation for which cavity modes can be built up), it was necessary to
take into account a high number of multipoles in order to converge both
the scattering cross section and the local field enhancement.  For
example, for a sphere radius of 60~nm, for reproducing the main wide
peak of the scattering cross section it was enough to retain multipoles
up to around $l=10$. However, we needed to take into account multipoles
up to the order of $l=70$ to arrive at convergence for the local field
enhancement and to reproduce the peak of the scattering cross section at
short sphere-surface distance associated with a higher resonant mode.
For the range of energies where tip-surface plasmon modes can be built
up, a simple model is inadequate to describe the system.

Cavity or gap modes have been reported in the context of light emission
in STM~\cite{berndt,nilius} and in samples consisting of metallic
nanoparticles placed very close to a metallic surface.~\cite{kume} In
both cases, the distance between the metallic tip or nanoparticle and
the metallic surface can be shorter than 1 nanometer.  For SNOM, little
has been done in connection with gap modes.  In the context of
apertureless SNOM, the approach curves have been experimentally studied
in detail by several
groups.~\cite{fischer,specht,kim,koglin,ferber,hillenbrand,andre}
A usual effect is the appearance of a peak or maximum of the signal when
tip approaches the sample, especially when both the tip and the sample
are made of the same metal.~\cite{ferber} This peak, which is sensitive
to the polarization and the dielectric functions of the tip and the
sample, can be interpreted as a manifestation of the fundamental cavity
mode discussed in this article.  In a recent article,~\cite{andre}
Andr\'e {\it et al.}\ have reported experimental results on apertureless
SNOM with a STM that might be related to the formation of cavity modes
between the tip and the surface.  Particularly, different behaviors of
the signal as a function of the tip-substrate distance have been
measured, with the appearance of minima and maxima as the tip is removed
from the substrate. There are some qualitative similarities with the
results shown in Fig.~\ref{fig:1.9eV}, supporting the claim that the
effects observed in Ref.~\onlinecite{andre} are associated with cavity
modes. Nevertheless, the experimental setup in Ref.~\onlinecite{andre}
is different than the one considered in the present work, therefore the
comparison should be established with caution.  

So far, we have considered sphere-surface distances much shorter than
the wavelength. For sphere-surface separations on the order or larger
than the wavelength of the incident wave, both the scattering
cross-section and the near-field enhancement oscillate as a function of
the tip-surface distance.  These undulations are due to the interference
of the incident wave and the wave reflected by the surface.  At this
range of sphere-surface distance, there is no significant coupling
between the sphere and the sample. 

\section{Conclusions}
\label{sec:conclusions}

By means of a detailed theoretical framework, which includes
retardation, higher multipoles of the tip, and multiple scattering
between the tip and the sample, we have studied the effects associated
with the tip-sample coupling in apertureless SNOM.  In particular, we
have analyzed the implications of energy shifts of the tip-induced
surface plasmons or cavity modes as the tip approaches the surface.
There is no exact correspondence between the local field enhancement and
the scattering cross section.  As a function of the tip-surface
distance, the scattering cross section present maxima related to the
cavity modes, while the local field enhancement keeps increasing as the
tip approaches the surface.  For a photon energy for which resonant
effects are negligible, both the local field enhancement and the
scattering cross section present maxima but at well separated values of
the tip-surface distance, in contrast to what is sometimes assumed. In
some cases, the maximum of the scattering cross section appears at
tip-surface distances of 15~nm while the maximum local-field enhancement
is at a tip-surface distance of approximately 2~nm. 

We have also studied the convergence of the results in terms of the
number of multipoles included in the calculation.  In general, one must
retain a high number of multipoles in order to arrive at convergence.
Only for energies for which no cavity modes are built up, a simple model
(retaining dipole and quadrupole modes of the sphere) for the
sphere-surface system can be appropriate for estimating the light
scattered off the system, but insufficient for the local field
enhancement.  When tip-surface plasmon modes are present, the simple
model is inadequate and a high number of multipoles should be taken into
account. 

Finally, we would like to stress that the present results are also
significant for surface-enhanced Raman scattering~\cite{nie} and the
emerging area of tip-enhanced Raman
spectroscopy.~\cite{stockle,anderson,hayazawa} The formation of cavity
modes between the tip and the sample and the associated field
enhancement can play a significant role.~\cite{mills}

\section*{Acknowledgments}
JAP gratefully acknowledges financial support from the European Union
through TMR network Electromagnetic Interactions in Tunneling (Contract
No. ERB-FMRX-CT98-0198). The research of PJ and SPA is further supported
by the Swedish Natural Science Research Council.  The work of PJ is also
supported by the SSF through the Nanometer Consortium at Lund
University.

\end{document}